\documentclass[12pt]{amsart}

\usepackage{latexsym}
\usepackage{graphicx}
\usepackage{cite}

\theoremstyle{plain}

\newtheorem{theo}{Theorem}
\newtheorem{corr}{Corollary}

\theoremstyle{definition}

\theoremstyle{remark}

\begin{document}

\thispagestyle{empty}

\date{\today}

\thanks{The research reported here
was performed while the authors were at MIT, Cambridge. It was
partially supported by the ``Communaut\'e francaise de Belgique -
Actions de Recherche Concert\'ees", by the EU HYCON Network of
Excellence (contract number FP6-IST-511368), by the Belgian
Programme on Interuniversity Attraction Poles initiated by the
Belgian Federal Science Policy Office, and by the DoD AFOSR URI
for ``Architectures for Secure and Robust Distributed
Infrastructures", F49620-01-1-0365 (led by Stanford University).
The scientific responsibility rests with its authors. Rapha\"el
Jungers is a FNRS fellow (Belgian Fund for Scientific Research).}

\title{Observable graphs}

\author{
Rapha\"el M. Jungers}
\address{Division of Applied Mathematics,
Universit\'e catholique de Louvain, 4 avenue Georges Lemaitre,
B-1348 Louvain-la-Neuve, Belgium} \email{blondel@inma.ucl.ac.be}

\author{
Vincent D. Blondel}
\address{Division of Applied Mathematics,
Universit\'e catholique de Louvain, 4 avenue Georges Lemaitre,
B-1348 Louvain-la-Neuve, Belgium} \email{jungers@inma.ucl.ac.be}


\maketitle
Abstract. {An edge-colored directed graph is \emph{observable} if
an agent that moves along its edges is able to determine his
position in the graph after a sufficiently long observation of the
edge colors.  When the agent is able to determine his position
only from time to time, the graph is said to be \emph{partly
observable}. Observability in graphs is desirable in situations
where autonomous agents are moving on a network and one wants to
localize them (or the agent wants to localize himself) with
limited information. In this paper, we completely characterize
observable and partly observable graphs and show how these
concepts relate to observable discrete event systems and to local
automata. Based on these characterizations, we provide polynomial
time algorithms to decide observability, to decide partial
observability, and to compute the minimal number of observations
necessary for finding the position of an agent. In particular we
prove that in the worst case this minimal number of observations
increases quadratically with the number of nodes in the graph.
From this it follows that it may be necessary for an agent to pass
through the same node several times before he is finally able to
determine his position in the graph. We then consider the more
difficult question of assigning colors to a graph so as to make it
observable and we
prove that two different versions of this problem are  NP-complete.}\\

\begin{section}{Introduction}
Consider an agent moving from node to node in a directed graph
whose edges are colored.  The agent knows the colored graph but
does not know his position in the graph. From the sequence of
colors he observes he wants to deduce his position.   We say that
an edge-colored directed graph\footnote{The property of being
observable is a property of directed graphs that have their edges
colored. For simplicity, we will talk in the sequel about
observable \emph{graphs} rather than observable \emph{edge-colored
directed graphs}.} is \emph{observable} if there is some
observation time length after which, whatever the color sequence
he observes, the agent is able to determine his position. Of
course, if all edges are of different colors, or if edges with
different end-nodes are of different colors, then the agent is
able to determine his position after just one observation. So the
interesting situation is when there are fewer colors than there
are nodes.
\begin{figure}
\begin{tabular}{cc}
\includegraphics[width=0.3\textwidth ]{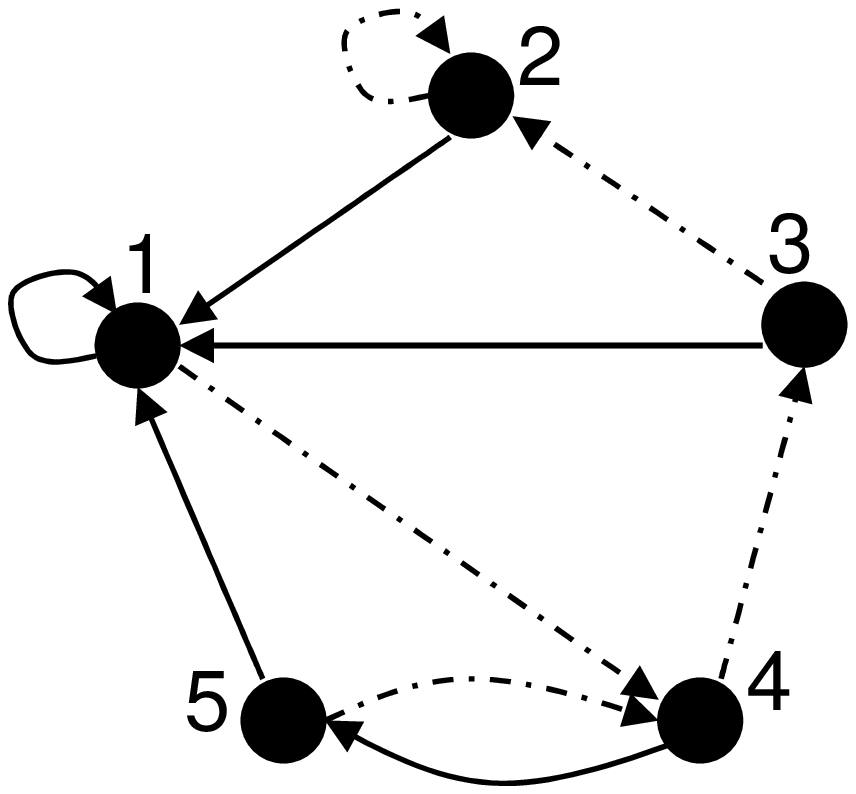} &
\includegraphics[width=0.3\textwidth ]{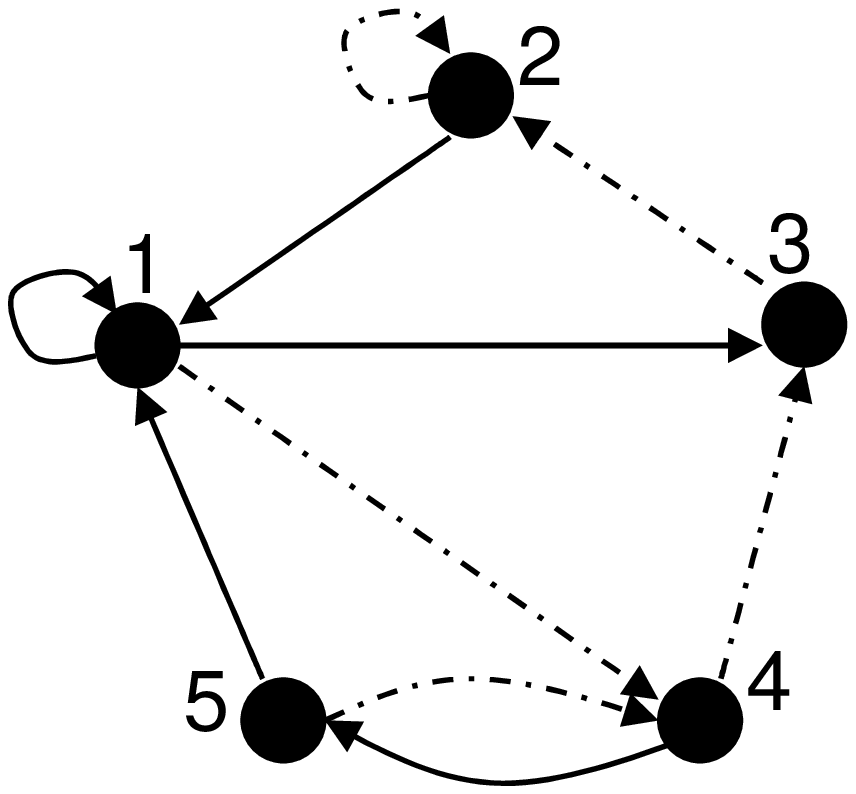}\\
(a) & (b)
\end{tabular}
\caption{An agent is moving along directed edges.
Edges are ``colored" solid (S) or dashed (D). The graph (a)
is observable because the observation of the last three colors suffices to determine the position of the
agent in the graph. The graph (b) is not observable; the sequence SSDSSD$\ldots$SSD can be obtained from a path ending
at node 2 or at node 4.}
\label{fig-ex-pos}
\end{figure}
Consider for instance the two graphs on Figure \ref{fig-ex-pos}.
The graphs differ only by the edge between the nodes 1 and 3. The
edges are colored with two ``colors": solid (S) and dashed (D). We
claim that the graph (a) is observable but that (b) is not. In
graph (a), if the observed color sequence is DDS then the agent is
at node 1, while if the sequence is SDD he is at node 3. Actually,
it follows from the results presented in this paper that the
observation of color sequences of length three always suffices to
determine the exact position of the agent in this graph. Consider
now the graph (b) and assume that the observed sequence is SSDSS;
after these observations are made, the agent may either be at node
1, or at node 3. There are two paths that produce the color
sequence SSDSS and these paths have different end-nodes. Sequences
of arbitrary large length and with the same property can be
constructed and so graph (b) is not observable.

There are of course very natural conditions for a graph to be
observable. A first condition is that no two edges of identical
colors may leave the same node. Indeed, if a node has two outgoing
edges with identical colors then an agent leaving that node and
observing that color will not be able to determine his next
position in the graph. So this is clearly a necessary condition.
Another condition is that the graph may not have two cycles with
identical color sequences, as for example the cycles $2-1-3-2$ and
$4-5-1-4$ in graph (b) of Figure \ref{fig-ex-pos}. If such cycles
are present in the graph, then an agent observing that repeated
particular color sequence is not able to determine if he is moving
on one or the other cycle. So, these two conditions are clearly
necessary conditions for a graph to be observable. In our first
theorem we prove that these two conditions together are also
sufficient.

In an observable graph there is some $T\geq 1$ for which an agent
is able to determine his position in the graph for all
observations of length $T$, and so he is also able to do so for
all subsequent times. Some graphs are not observable in this way
but in a weaker sense. Consider for example the graph (a) of
Figure \ref{fig-weak}. This graph is not observable because there
are two dashed edges that leave the top node and so, whenever the
agent passes through that node the next observed color is D and
the position of the agent is then uncertain. This graph is
nevertheless \emph{partly observable} because, even though the
agent is not able to determine his position at all time, there is
a finite window length $T$ such that the
agent is able to determine his position at least once every $T$
observations. Indeed, after observing the sequence SD in this
graph (and this sequence occurs in every observation sequence of
length 4), the next observation is either S or D. If it is S, the
agent is at the bottom right node;  if it is D, he is at the
bottom left node. In both cases the next observed color is S and
the agent is then at the top node. So the agent is able to
determine his position at least once every five observations. In
an observable graph, an agent is able to determine his position at
all times beyond a certain limit; in a partly observable graph the
agent is able to determine his position infinitely often (for
formal definitions, see below).

Notice that in the graph (a) of Figure \ref{fig-weak}, the agent
is able to reconstruct its entire trajectory \emph{a posteriori},
except maybe for its last position. This is however not the case
for all partly observable graphs. Consider for example the star
graph (b) of Figure \ref{fig-weak}. The color sequences observed
in this graph are SDSDS$\ldots$ or DSDSDS$\dots$. When the last
observed color is D, the agent is at the central node. When the
last observed color is S, the agent only knows that he is at one
of the extreme nodes. Hence this graph is partly observable but in
this case, contrary to what we have with graph (a), the entire
trajectory cannot be reconstructed a posteriori.
\begin{figure}
\centering
\begin{tabular}{cc}
\includegraphics[width=0.3\textwidth ]{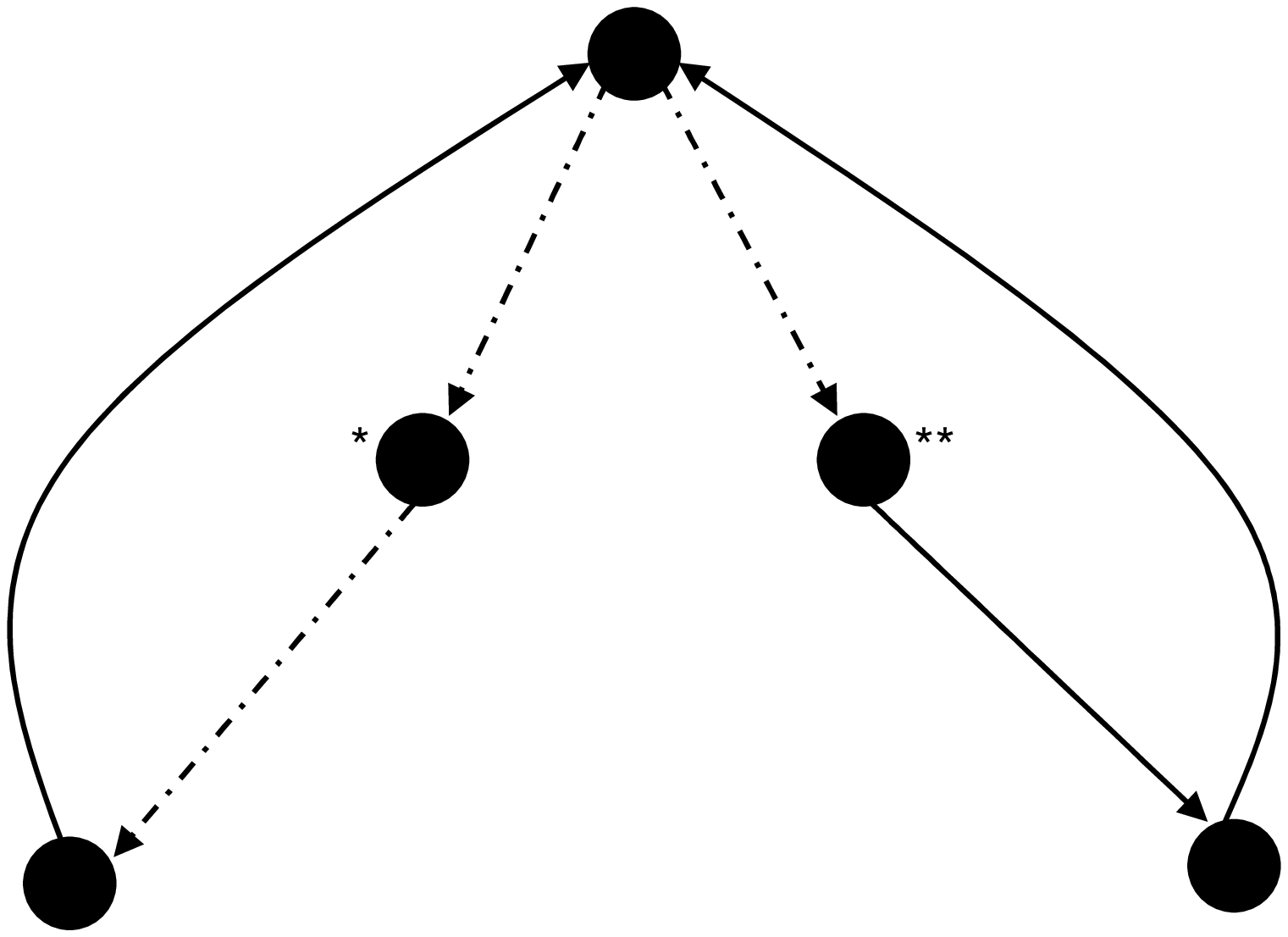} &
\includegraphics[width=0.3\textwidth ]{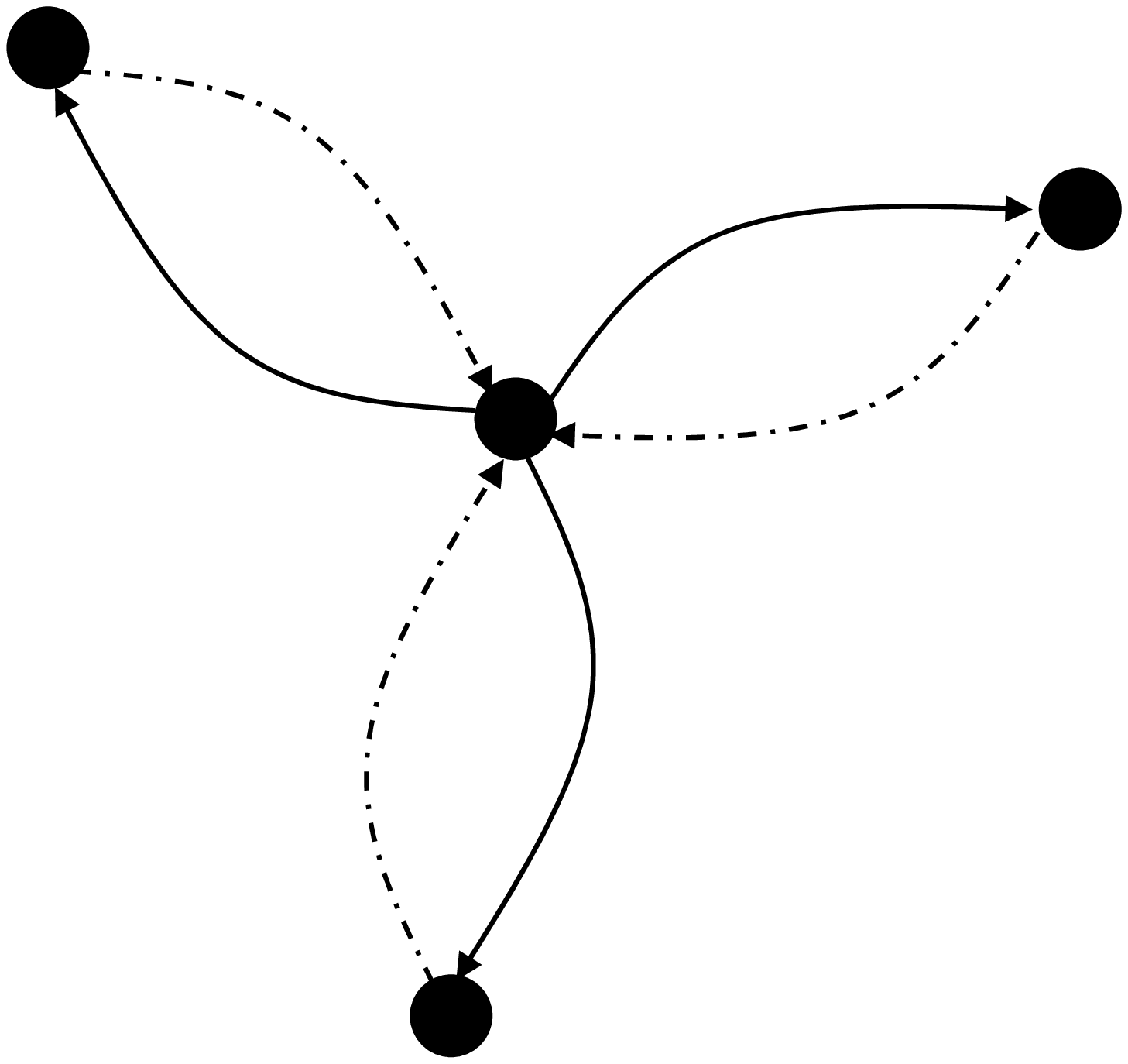}\\
(a) & (b)
\end{tabular}

\caption{Both graphs are partly observable but none of them is observable.
In graph (a), complete trajectories can
always be reconstructed a posteriori; which is not possible in
graph (b).} \label{fig-weak}
\end{figure}

In this paper, we prove a number of results related to observable
and partly observable graphs. We first prove that the conditions
described above for observability are indeed sufficient. From the
proof of this result it follows that in an observable graph an
agent can determine his position in the graph after an observation
of length \emph{at most $n^2$}, where $n$ is the number of nodes.
This quadratic increase of the observation length cannot be
avoided: we provide a family of graphs for which $\Theta(n^2)$
observations are necessary. Based on some of these properties we
also provide polynomial-time algorithms for checking if a graph is
observable or partly observable. We then consider the question of
assigning colors to the edges or nodes of a directed graph so as
to make it observable and we prove that the problem of finding the
minimal number of colors  is NP-complete.

The concept of observable graphs is related to a number of
concepts in graph and automata theory, control theory and Markov
models which we now describe.

Observable graphs as defined here are actually a particular case
of local automata. A finite state automaton is said to be
$(d,k)$-\emph{local} (with $d \leq k$) if any two paths of length
$k$ and identical color sequences pass through the same state at
step $d$. Hence an observable graph is a $(k,k)$-local automaton.
It is shown in \cite{Beal93} how to recognize local automata in
polynomial time but the motivation in that context is very
different from ours and little attention is given there at the
particular values of the integer $d$ and $k$. In particular, the
algorithm provided in \cite{Beal93} does not allow to recognize
$(k,k)$-local automata.

Related to the notion of observable graphs, Crespi et al.
\cite{Crespi05} have recently introduced the concept of
\emph{trackable graph}. An edge-colored directed graph is said to
be \emph{trackable} if the maximum number of trajectories
compatible with a color sequence of length $k$ grows
subexponentially with $k$. So, in the context of trackability, one
cares about the total number of compatible trajectories, but not
about the position of the agent in the graph. It has recently been
proved that the problem of determining if a graph is trackable can
be solved in polynomial time
 \cite{JungersProtasovBlondel06}; see also \cite{Crespi05}.
 It is clear that an observable network is trackable, but the converse is not true in general. For example the graph on
 Figure \ref{fig-weak} (a) is trackable
 but not observable. Notice also that, as shown with the graph on Figure \ref{fig-weak} (b),
  partly observable graphs do not need to be trackable.

Observable graphs are also related to Discrete Event Systems
(DES). More precisely, our notion of \emph{partly observable}
graph is similar to what \"Oszveren and Willsky call observable
DES \cite{OzverenWillsky90}. These authors define  DES as  colored
graphs, except for the fact that they allow  transitions to be
unobservable: some edges have no colors. From a colored graph with
unobservable transitions we can easily construct an equivalent
fully colored graph by removing all unobservable transitions and
adding an edge $(h, j)$ of color $c$ whenever there is an edge
$(h, i)$ of color $c$ and an unobservable transition (or a
sequence of unobservable transitions) between $i$ and $j$. Our
results are therefore applicable to observable DES as defined in
\cite{OzverenWillsky90}.

Finally, the results presented in this paper can also be
interpreted in the context of \emph{Hidden Markov Processes} (HMP)
\cite{Rabiner90,EphraimM02}.  More precisely, the graphs we
consider can be seen as \emph{finite alphabet HMPs} (also called
\emph{aggregated Markov processes}), except that no values
different from 0 and $1$ are given for the transition
probabilities. In our context, a transition is either allowed or
it is not; we do not associate transition probabilities. Also, we
can assign to a given transition several possible colors, but
again, without considering their respective probabilities.

The remainder of this paper is organized as follows: in Section
\ref{section-criterion} we formalize the notions presented above
and give necessary and sufficient conditions for observability.
Section \ref{section-algo} deals with algorithmic aspects: we show
there how to check the conditions derived in Section
\ref{section-criterion} in polynomial time. In Section
\ref{section-np} we prove NP-completeness results for the design
of observable and partly observable graphs. Finally, in Section
\ref{section-conc} we conclude and describe some open questions.

\end{section}
\begin{section}{Characterizing observability}\label{section-criterion}
Let $G=(V,E)$ be a graph and $C$  a set of colors.  To every edge
$e \in E$ we associate one (or more) color from $C$.  A word $w$
on $C$ is the concatenation $w_1\dots w_T$ of symbols taken from
$C$; the length $\vert w \vert$ of $w$ is the number of its
symbols. A subword $w_{[i,j]}:1\leq i\leq j \leq \vert w \vert$ of
$w=w_1\dots w_T$ is the concatenation of the symbols $w_i\dots
w_j$. We say that a path is allowed by a word $w$  if for all $i$
the $i$th edge of the path has color $w_i$.  Finally, for a word
$w$ and a set $S\subset V$, we denote by $\delta_w(S)$ the set of
nodes $q$ for which there exists a path allowed by $w$ beginning
in a state in $S$ and ending in $q$; $\delta_w(V)$ is the set of
nodes that can be reached from a node in $V$ with the color
sequence $w$.

A graph is \emph{observable} if there exists an integer $T$ such
that for all words $w$ of length $\vert w\vert \geq T$, $\vert
\delta_w(V)\vert \leq 1$. In other words there is at most one
possible end-node after a path of length $T$ or more. We shall
also say that a graph is \emph{partly observable} if there exists
an integer $T$ such that for all word $w$ with $\vert w\vert \geq
T$, there
exists a $1 \leq t\leq T$ such that $\vert \delta_{w_{[1,t]}}(V)\vert \leq 1$.\\
The theorem below characterizes observable graphs. For a graph to
be observable no color sequence $w$ can allow two \emph{separated}
cycles, that is, two cycles $\pi,\pi':[0,|w|]\rightarrow V$ such
that $\pi(i)\neq\pi'(i), \forall i$ and the edges
$(\pi(i),\pi(i+1))$ and $(\pi'(i), \pi'(i+1)$ have identical
colors for every $i$. For instance, the graph in Figure
\ref{fig-ex-pos} (b) is not observable because the color sequence
``solid-dashed-solid'' allows both the cycle $1-3-2-1$ and the
cycle $5-1-4-5$.  Another condition for a graph to be observable
is that no node may have two outgoing edges of the same color.
This last condition only applies to the nodes of the graph that
are \emph{asymptotically reachable}, i.e., that can be reached by
paths of arbitrary long lengths. In a strongly connected graph,
all nodes are asymptotically reachable. For graphs that are not
strongly connected, asymptotically reachable nodes are easy to
identify in the decomposition of the graph in strongly connected
components. We now prove that these two conditions together are
not only necessary but also sufficient:
\begin{theo}\label{theo-pos}
An edge-colored graph is observable if and only if no color
sequence allows two separated cycles, and no asymptotically
reachable node of the graph has two outgoing edges that have
identical colors.
\end{theo}
\begin{proof}
$\Rightarrow:$ We prove this part by contraposition. Suppose that
the first condition is violated: if we have a word $w$ allowing
two separated cycles, then $ww \cdots w$  is an arbitrarily long
word that allows two separated paths.  Suppose now that the second
condition is violated and let $v$ be an asymptotically reachable
node. For any $T$, we can find a path of length $T-1$ ending in
$v$ and $v$ has two outgoing edges of the same color. Let $w$ be a
color sequence allowing this path, and $c$ the color of the two
edges. We have then an arbitrarily long word $wc$ such that $\vert
\delta_{wc}(V)\vert\geq 2$.

 $\Leftarrow:$ Let us define $T_1$ such
that every path of length larger than $T_1$ has its last node
asymptotically reachable.  We consider two arbitrary compatible
paths, that is,  two paths allowed by the same color sequence. We
will show that they intersect between $t=T_1$ and $t=T_1+n'^2$,
with $n'$ the number of asymptotically reachable nodes.  This will
establish observability of the graph, since two paths cannot split
once they intersect (recall that no node has two outgoing edges of
the same color), and so actually \emph{all} compatible paths pass
through the same node after $t=T_1+n'^2$ steps. Suppose by
contradiction that the two paths $\pi, \pi'$ allowed by the same
color sequence do not intersect during the last $n'^2$ steps, when
they only visit asymptotically reachable nodes. Then by the
pigeonhole principle there are two instants $t_1,t_2$ such that
$(\pi(t_1), \pi'(t_1))=(\pi(t_2), \pi'(t_2))$, and the first
condition is violated: $w_{[t_1,t_2]}$ is a sequence that allows
two separated cycles.
\end{proof}

One could think that the first condition in Theorem \ref{theo-pos}
suffices for the graph to be partly observable, but this is not
the case. Consider indeed the graph in Figure \ref{fig-weak-real}.
For notational simplicity we have put colors on the nodes rather
than on the edges: one can interpret these node colors as being
given to the edges incoming in the corresponding node.  This graph
satisfies the first condition but it is not partly observable.
Indeed, the observed sequence $RGGRGG\dots$ has no subsequence
allowing only one ending node.

Actually, the first condition implies a weaker notion of
observability: A graph is \emph{partly a-posteriori observable} if
for any sufficiently long observation, it is possible \emph{a
posteriori} (that is, knowing the whole observation) to determine the state of the agent at at least one
previous instant. We do not develop this concept any further in
this paper but it should be clear that all the proofs that we
provide in this and in the next section can easily be extended to
cover that case as well.

The graph (a) of Figure \ref{fig-ex-pos} is observable and the
agent can be localized after an observation of length at most 3.
The above proof provides an upper bound on the observation length
that is needed to localize an agent in an observable graph.

\begin{figure}
\centering
\includegraphics[width=0.5\textwidth ]{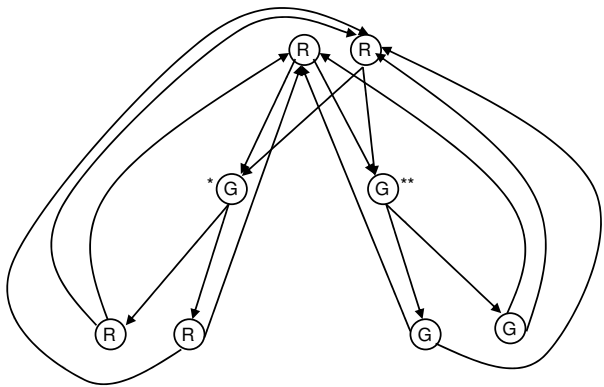}
\caption{A graph that is not partly observable, even though no color sequence allows two separated cycles.  The colors on nodes can be understood as assigned to the edges incoming in the corresponding nodes.}
\label{fig-weak-real}
\end{figure}

\begin{corr}\label{cor-nsquare}
In an observable graph, $n^2-n$ observations suffice to localize
an agent ($n$ is the number of nodes in the graph).
\end{corr}
\begin{proof}
Let us suppose by contradiction that there exists a color sequence
of length $T>n^2-n$ that allows two paths ending in different
nodes.  There are $n^2-n$ different couples of nodes $(v_1,v_2):
v_1 \neq v_2$ and one can follow a similar reasoning as in the
previous proof to show that the graph is actually not observable.
\end{proof}

One could have expected the maximal value of $T$ to be $n$, and
one may wonder whether the bound $n^2-n$ in the theorem is tight.
We give in Section \ref{section-algo} a family of graphs for which
the value of $T$ is $n(n-1)/2-1$; the quadratic increase can
therefore not be avoided.

\end{section}

\begin{section}{Verifying observability}\label{section-algo}
In this section, we consider two algorithmic problems. The first
problem is that of reconstructing the possible positions of an
agent in a graph for a given sequence of color observations. That
problem is easy; we describe a simple solution to it and give an
example that illustrates that the bound in Corollary
\ref{cor-nsquare} is tight. The second problem is that of deciding
observability.

Let us consider the first problem: we are given a color sequence
and we would like to identify the nodes that are compatible with
the observed sequence of colors. A simple algebraic solution is as
follows. To every color $c$, there is an associated graph $G_c$
for which we can construct the corresponding adjacency matrix
$A_c$. To a color sequence $w=w_1, \ldots, w_{\vert w\vert}$ we
then associate $A_w$, the corresponding product of matrices
$A_w=A_{w_1}\dots A_{w_{\vert w\vert}}$. It is easy to verify that
the $(i,j)$th entry of $A_{w}$ is equal to the number of paths
from $i$ to $j$ allowed by $w$. The set $\delta_w(S)$ is therefore
obtained by restricting the matrix $A_w$ to the lines
corresponding to elements in $S$, and by taking the indices of the
nonzero columns. This simple algorithm is actually nothing else
than the well-known Viterbi algorithm for Hidden Markov Processes
\cite{Viterbi67,Rabiner90}, except that in our context the
probabilities are all equal to zero or one.

As an application, let us show that the quadratic dependance  for
the bound in Corollary \ref{cor-nsquare} cannot be improved. We
describe a family of colored graphs with $n$ nodes by giving the
adjacency matrices corresponding to the different colors. Our set
contains $2(n-2)$ matrices.  For every $1\leq i \leq n-2$ we
construct two matrices: the first one, $A_i$, has $n-i$ entries
equal to one, which are the entries $(i+1,i+2),\dots,(n-1,n),$ and
$(i,i)$; the second matrix, $B_i$, has two entries equal to one:
the entries $(i,i+1)$ and $(n,i+2)$. A colored graph defined by
this set of matrices is observable (the reader can verify this by
using the algorithm given below); but on the other hand the color
sequence $A_1^{n-2}B_1A_2^{n-3}B_2\dots A_{n-2}B_{n-2}$, of length
$n(n-1)/2-1$, allows paths that have different end nodes.  This is
straightforward to check by computing the corresponding product of
matrices and check that it has two columns with nonzero elements.

We now turn to the main question of this section: how can one
determine efficiently if a graph is observable? We have the
following result.

\begin{theo}
The problems of determining whether a colored graph is observable, or partly observable, are solvable in
$O(n^4m)$, where $n$ is the number of nodes, and $m$ is the number of colors.
\end{theo}
\begin{proof}
The algorithm we propose uses the conditions given in Theorem
\ref{theo-pos}. We first check the condition that no color
sequence allows two separated cycles. In order to do that, we
construct an auxiliary directed graph that we will denote $G^2$
and  whose nodes are couples of distinct nodes in $G$:
$\{(v_1,v_2)\in V^2:v_1\neq v_2\}$. In $G^2$ there is an edge from
$(v_1,v_2)$ to $(v_1',v_2')$ if there is a color that allows both
edges $(v_1,v_1')$ and $(v_2,v_2')$.  This graph can be
constructed in $O(n^4m)$ operations, since there are less than
$n^4$ edges in $G^2$ and for each of these edges we have to check
$m$ colors. It is possible to design two separated cycles in $G$
allowed by the same color sequence if and only if there is
\emph{one} cycle in $G^2$. This we can test by a simple breadth
first search. Now, we can check easily if an asymptotically
reachable node has two outgoing edges of the same color, and
determine whether the network is observable.

 We now would like to
test whether the graph is partly observable; in order to do that,
we construct a new auxiliary directed graph $\tilde{G}^2$, and
once again we check the existence of cycles in this graph.  An
easy way to obtain this new graph is to take $G^2$, and to add
some edges: for every couple of nodes $((v_1,v_2),(v_1',v_2'))$,
we allow an edge if there is a color that allows one edge from
$v_1$ \emph{or} $v_2$ to $v_1'$ and one edge from $v_1$ \emph{or}
$v_2$ to $v_2'$.  The graph $\tilde{G}^2$ can also be constructed
in $O(n^4m)$ operations.  We claim that the graph is partly
observable if and only if this auxiliary graph is acyclic. Indeed,
any path in the auxiliary graph ending in node, say, $(v_1,v_2)$
corresponds to a color sequence that allows paths ending in each
of the nodes $v_1$ and $v_2$.  If the auxiliary graph contains a
cycle, consider the path travelling along the cycle during $T$
steps; it corresponds to a word of length $T$ such that for any
$t\leq T$, the subword obtained by cutting the word at the $t$th
character allows two paths ending in different nodes, and the
definition of partial observability is violated. On the other
hand, if the graph is not partly observable, there exist
arbitrarily long sequences such that all beginning subsequence
allows at least two paths ending in different nodes. This implies
that $\tilde{G}^2$ has a cycle. Indeed, let us consider such a
color sequence $\omega$ of length $T \geq n^2$. There are two
paths $\pi,\pi_1: [0,T]\rightarrow V$ allowed by $\omega$ and
ending in two different nodes: $\pi(T)\neq \pi_1(T)$. If
$\pi(T-1)\neq \pi_1(T-1),$ then there is an edge in $\tilde{G}^2$
from $(\pi(T-1), \pi_1(T-1))$ to $(\pi(T),\pi_1(T))$.  If
$\pi(T-1) = \pi_1(T-1),$ then by construction of $\omega$ there
exists a path $\pi_2$ allowed by $\omega_{[1\dots T-1]}$ that ends
in a node $v_{T-1}\neq \pi(T-1).$  Thus we can iteratively define
nodes $v_i: i\in [1,T]$ of decreasing indices $i$ such that there
is a path in $\tilde{G}^2$ from $(v_{i-1}, \pi(i-1))$ to $(v_{i},
\pi(i)).$ Finally this path is of length $T$ which is more than
the number of nodes in $\tilde{G}^2,$ so this graph is cyclic.
Since $\tilde{G}^2$ can be constructed in $O(n^4m)$, the proof is
complete.
\end{proof}
This algorithm also allows us to compute the time necessary before localization of the agent:
\begin{theo}
In an observable graph, it is possible to compute in polynomial time the minimal $T$
such that any observation of length $T$ allows to localize the agent. The same is true
for partial observability.
\end{theo}
\begin{proof}
Let us consider the graphs $G^2$ and $\tilde{G}^2$ in the above
proof. The existence of an observation of length $T$ allowing two
different last nodes (respectively, two paths with at least two
different allowed nodes at every time) is equivalent to the
existence of a path of length $T$ in $G^2$ (respectively,
$\tilde{G}^2$). Since the auxiliary graphs are directed and
acyclic, it is possible to find a longest path in polynomial time.
\end{proof}

\end{section}

\begin{section}{Design of observable graphs is NP-complete}\label{section-np}

In this section, we prove NP-completeness results for the problem
of designing observable graphs. The problem is this: we are given
an (uncolored) graph and would like to color it with as few colors
as possible so as to make it observable. Unless the graph is
trivial, one color never suffices. On the other hand, if we have
as many colors as there are nodes the problem becomes trivial. A
simple question is thus: ``What is the minimal number of colors
that are needed to make the graph observable''? We show that this
problem is NP-complete in two different situations: If we color
the nodes and want to make the graph observable, or if we color
the edges and want to obtain a partly observable graph. We haven't
been able to derive similar results for the other two combinations
of node/edges coloring and observability/partial observability. In
particular, we leave open the natural question of finding the
minimal number of edge colors to make a graph observable. Even
though the statements of our two results are similar, the proofs
are different. In particular, we use reductions from two different
NP-complete problems (3-COLORABILITY and MONOCHROMATIC-TRIANGLE).

\begin{theo}
The problem of determining, for a given graph $G$ and integer $k$,
if $G$ can be made  observable by coloring its nodes with $k$
colors is NP-complete.
\end{theo}
\begin{proof}
We have already shown how to check in polynomial time if a colored
graph is observable. So the problem is in NP.  In the rest of the
proof we provide a polynomial-time reduction from 3-COLORABILITY
to the problem of determining if a graph can be made observable by
coloring its nodes with 3 colors. This will then establish the
proof since 3-COLORABILITY is known to be NP-complete
\cite{GJ-computers-igt}.

In 3-COLORABILITY we are given an undirected graph $G$ and the
problem is to color the nodes with at most three different colors
so that no two adjacent nodes have the same color. For the
reduction, consider an undirected graph $G=(V,E)$.  From $G$ we
construct a directed graph $G'=(V',E')$ such that $G'$ can be made
observable by node coloring with 3 colors if and only if $G$ is
3-colorable. Figure \ref{fig-np-demo} illustrates our
construction. On the left hand side we have a $3$-colorable
undirected graph. The corresponding directed graph $G'$ is
represented on the right hand side. The graph $G'$ can be made
observable with three colors.
\begin{figure}
 \centering
\includegraphics[width=0.6\textwidth ]{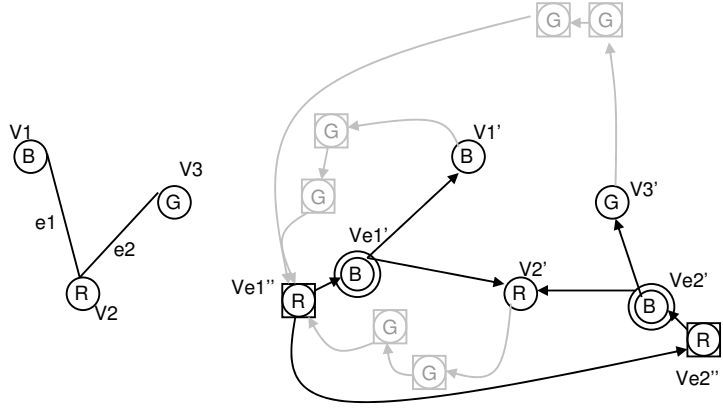}
 \caption{Construction of the graph $G'$ and its coloration}
\label{fig-np-demo}
 \end{figure}
Here is how we construct $G'$ from $G$: For every node $v$ in $V$,
we define a corresponding node $v'$ in $V'$ and call these nodes
``real nodes''. For every edge $e=(v_1,v_2)$ in $E$, we define a
node $v'_e$ in $V'$ and two outgoing edges $(v'_e, v_1')$ and
$(v'_e, v'_2)$. The nodes $v'_e$ appear with double circles on
Figure \ref{fig-np-demo}. The main idea of the reduction is as
follows: If $G'$ is observable by node coloring, then $v_1'$ and
$v_2'$ cannot possibly have the same color since there are edges
pointing from $v'_e$ to both $v_1'$ and $v_2'$, and so the
corresponding nodes in $G$ do indeed have different colors.

The graph $G'$ we have constructed so far is rather simple and
does not contain any cycle; in the remainder of our construction
we add nodes and edges to $G'$ so as to make it strongly connected
but in such a way that the graph $G'$ remains observable. For
every node $v_e'$, we add a new node $v_e''$ and an edge
$(v_e'',v_e')$. The nodes $v_e''$ are represented by a square on
Figure \ref{fig-np-demo}. We then order the edges in $G$:
$E=\{e_1,\dots,e_s\}$, and we add edges
$(v_{e_i}'',v_{e_{i+1}}'')$ in $G'$ for every $1\leq i \leq s-1$.
Finally, for every real node $v'$ in $G'$ we add two nodes, with
an edge from $v'$ to the first one, an edge from the first one to
the second one, and an edge from the second one to $v_{e_{1}}''$.
These nodes and edges are represented in light grey on Figure
\ref{fig-np-demo}. The graph $G'$ is now strongly connected and we
claim that it can be made observable by node coloring with three
colors if and only if $G$ is $3$-colorable. Let us establish this
claim.

If $G'$ can be made observable by node coloring with three colors
then clearly $G$ is 3-colorable. Indeed, whenever two nodes in $G$
are connected by an edge, their corresponding real nodes in $G'$
are being pointed by edges emanating from the same node and so
they need to have different colors. So this part is easy.

Assume now that $G$ is 3-colorable and let its nodes be colored in
red (R), blue (B) or green (G). We need to show that the graph
$G'$ can be made observable by coloring its nodes with three
colors. We choose the following colors: every real node is colored
by its corresponding color in $G$, the nodes $v_e'$ are all
colored in blue and the nodes $v_e''$ are all colored in red.
Finally, the remaining nodes (in light grey on the figure) are
colored in green. This colored graph is clearly observable.
Indeed, any sufficiently long directed path (in particular, any
path whose length is larger than the total number of nodes in
$G'$) will eventually reach a succession of two green nodes
followed by a red node. At the end of such a sequence, the agent
is at node $v''_{e_1}$ and his position in the graph can then be
observed for all subsequent steps since $G'$ has no two outgoing
edges from a node leading to nodes of identical colors. Thus $G'$
is observable and this concludes the proof.

\end{proof}

The next result also deals with observability but the reduction is
quite different.

\begin{theo}
The problem of determining, for a given graph $G$ and integer $k$,
if $G$ can be made partly observable by coloring its edges with
$k$ colors is NP-complete.
\end{theo}

\begin{proof}

\begin{figure}
 \centering
\includegraphics[width=0.8\textwidth]{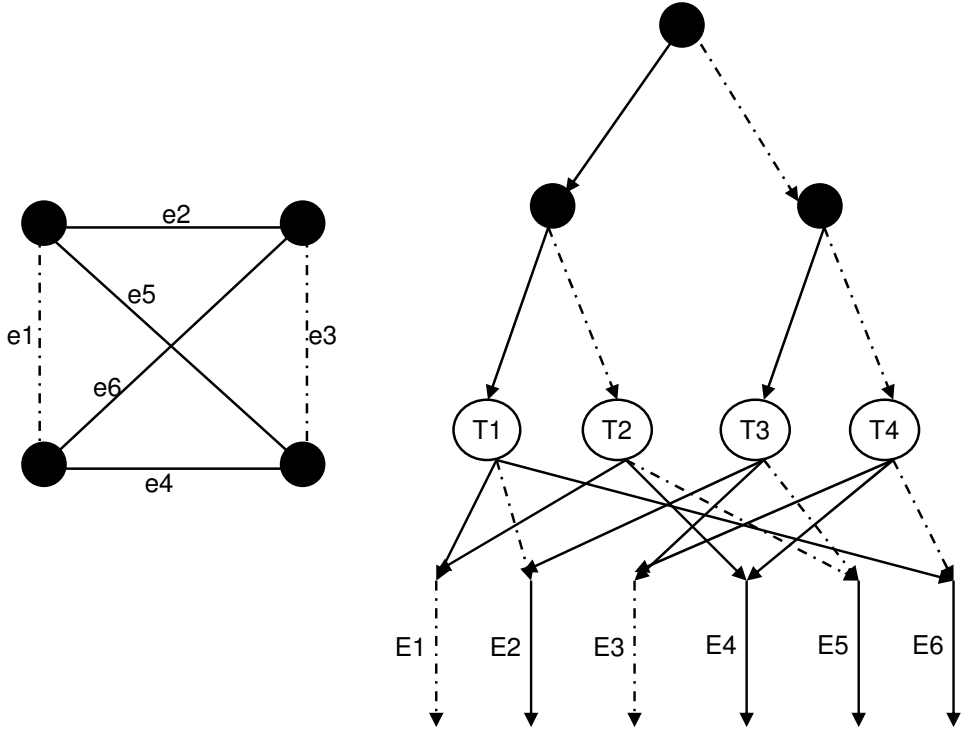}

 \caption{The main building block for the reduction.}
\label{fig-np-edge-demoter}
 \end{figure}

\begin{figure}
 \centering
\includegraphics[width=0.8\textwidth]{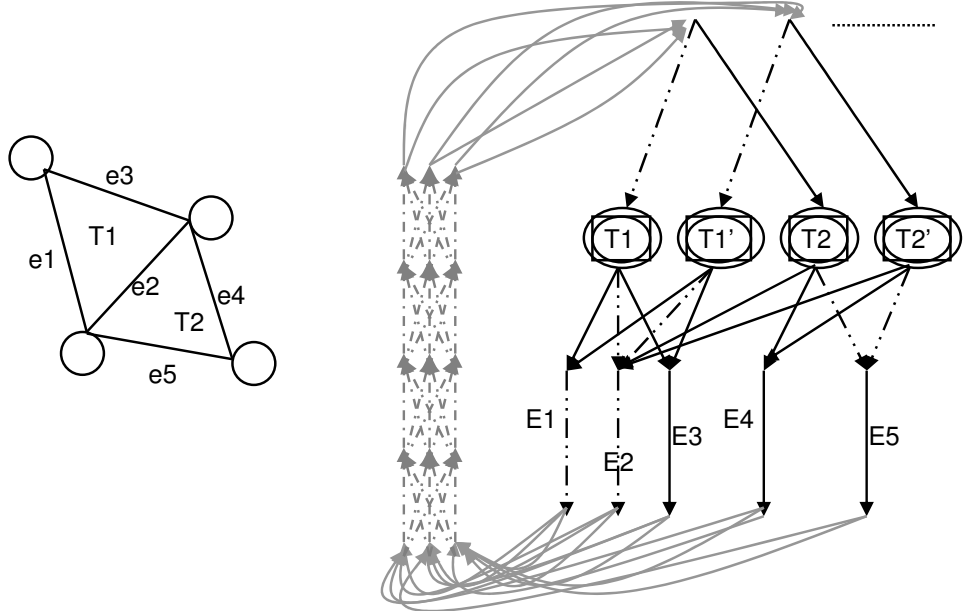}

 \caption{The complete reduction.}
\label{fig-np-edge-demobis}
 \end{figure}

We have shown in Section \ref{section-algo} how to check in
polynomial time if a colored graph is partly observable. Hence the
problem is in P and to complete the proof it suffices to exhibit a
polynomial time reduction from some NP-complete problem. Our proof
proceeds by reduction from MONOCHROMATIC-TRIANGLE. In
MONOCHROMATIC-TRIANGLE we are given an undirected graph $G$ and we
are asked to assign one of two possible colors to every edge of
$G$ so that no triangle in $G$ (a triangle in a graph is a clique
of size three) has its three edges identically colored. A graph
for which this is possible is said to be triangle-colorable and
the problem of determining if a given graph is triangle-colorable
is known to be NP-complete \cite{GJ-computers-igt}.

Our proof proceeds as follows:  From an undirected graph $G$ we
construct a directed graph $\tilde{G}$ that can be made partly
observable with two colors if and only if $G$ is
triangle-colorable. Thus we prove the somewhat stronger result
that the problem of determining if a graph can be made partly
observable with \emph{two} colors, is NP-complete.

As is often the case for NP-hardness proofs the idea of the
reduction is simple but the construction of the reduction is
somewhat involved. For clarity we construct $\tilde{G}$ in two
steps. The first step is illustrated on Figure
\ref{fig-np-edge-demoter}. In this first step we construct an
acyclic graph $G'$ that will be used in the second step. The
encoding of MONOCHROMATIC-TRIANGLE is done at this step. The
second step is illustrated on Figure \ref{fig-np-edge-demobis} and
uses several replica of parts of the construction given in the
first step. This second step is needed in order to make the final
graph $\tilde{G}$ connected.

Let us now consider the first step. The left hand side of Figure
\ref{fig-np-edge-demoter} represents an  undirected graph $G$. The
edges of $G$ are colored but we do not pay attention to the colors
at this stage. At the right hand side of the figure is the graph
$G'$ constructed from $G$. In order to construct $G'$, we order
the edges of $G$ and for every edge $e_i$ in $G$ we define two
nodes in $G'$ and a directed edge $E_i$ between them; these edges
have no nodes in common and we refer to them as ``real edges"
because they represent edges in the original graph $G$. Next, we
identify the triangles $\{e_q,e_r,e_s:q<r<s\}$ in $G$, we order
them (for example, according to the lexicographic order), we add a
node $T_i$ in $G'$ for every triangle in $G$, and we add in $G'$
edges from the node $T_i$ to all three real edges it is composed
of, as illustrated on the second level of the graph on Figure
\ref{fig-np-edge-demoter}. Thus the out-degrees of the nodes $T_i$
are all equal to three. Finally, we construct a binary tree whose
leaves are the nodes $T_i$. In the example represented on the
figure, the graph $G$ has
 four triangles and so the tree has four leaves $T_i$ and has depth 2. (In
general, the tree has depth equal to $N=\lceil \log_2{S} \rceil$
where $S$ is the number of triangles in $G$.) The reduction will
be based on the following observation: Let the edges of $G'$ be
colored and let us consider the sequences of colors observed on
paths from the root of the tree to the bottom nodes. We claim that
the edges of $G'$ can be colored with two colors so that all color
sequences observed on these paths are different if and only if $G$
can be triangle-colored. We now establish this claim. Assume first
that $G$ can not be triangle-colored and let $T_i$ be a node
corresponding to a triangle whose edges are identically colored.
There are three outgoing edges from $T_i$ and so two of them must
have identical colors. Since these edges lead to real edges of
identical colors, $G'$ has two paths that have identical color
sequences. Assume now that the graph $G$ can be triangle-colored.
We color the real edges of $G'$ with their corresponding colors in
$G$ and we color the edges of the tree so that paths from root to
leaves define different color sequences.  It remains to color the
outgoing edges from the nodes $T_i$. Every $T_i$ has three
outgoing edges $u_i, v_i, w_i$ and these edges lead to real edges
that are not all identically colored. Since there are only two
colors, two of the three edges must have identical colors. Let
$u_i$ and $v_i$ be the outgoing edges from $T_i$ that lead to the
two identically colored real edges; we give to $u_i$ and $v_i$
different colors and choose an arbitrary color for the third
outgoing edge $w_i$. When coloring the edges of $G'$ in this way,
all paths from the root to the bottom nodes define different color
sequences and so the claim is established.

We now describe the second step of the construction. This step is
illustrated on Figure \ref{fig-np-edge-demobis} where we show a
graph $G$ (on the left hand side) and the corresponding graph
$\tilde G$ (on the right hand side) constructed from $G$. Notice
that the graph $G$ we consider here is different from the graph
used in Figure \ref{fig-np-edge-demoter}. The construction of
$\tilde G$ consists in repeating a number of times the tree
structure described in the first step and then making the graph
connected by connecting the end nodes of the real edges to the
roots of the trees through a sequence of edges, as represented on
the figure. For the sake of clarity we have represented only two
copies of the tree on the figure but the complete construction has
$2S+1$ such copies (in the example of the figure, 5 copies); the
reason for having $2S+1$ copies will appear in an argument given
below. In the sequence of edges that make the graph connected
there are three nodes at every level and there are $N+3$ levels
where $N$ is the depth of the trees. The aim of this last
construction  is to make us able, by appropriately coloring the
edges, to determine when an agent reaches the root of a tree, but
without making us able to determine what particular root the agent
has reached.

We are now able to conclude the proof. We want to establish that
$\tilde G$ can be made partly observable with two colors iff $G$
is triangle-colorable.

Assume first that $G$ is triangle-colorable and let the coloration
be given with the colors dashed (D) and Solid (S). We color
$\tilde G$ as follows: For the real edges and for the edges in the
trees we choose the colors described in step 1, for the edges
leaving the end nodes of the real edges and for the incoming edges
to the roots we choose S, finally, all the other connecting edges
are colored by D. We claim that the graph $\tilde G$ so colored is
partly observable. In order to establish our claim consider an
agent moving in $\tilde G$. After some time the agent will hit a
long sequence of $N+3$ connecting edges colored by D. The agent
will know that he is at a root as soon as he observes a S. He will
however not be able to determine what particular root he is at. We
then use the argument presented in step 1 to conclude that, when
the agent subsequently arrives at one of the end node of the real
edges, he knows exactly at what node he is. Therefore the colored
graph $\tilde G$ is partly observable because, whenever the agent
reaches the end node of a real edge, he knows where he is.

We now establish the other direction: If $G$ is not
triangle-colorable then $\tilde G$ cannot be made partly
observable with two colors. We establish this by proving that when
$G$ is not triangle-colorable, then there exist two separated
cycles in $\tilde G$ that have identical color sequences. Since
the color sequences are identical and the cycles are separated an
agent observing that particular repeated color sequence can never
determine on what cycle he is and the graph is not partly
observable. Let us assume that some coloration for $\tilde G$ has
been chosen. Since $G$ is not triangle-colorable, some triangles
have all their edges identically colored. Choose one such triangle
and consider in every tree in $\tilde G$ the path that leads from
the root to the node corresponding to that triangle in the tree.
Any such path defines a color sequence and there are at most $S$
such sequences. Since there are $2S+1\geq S+1$ trees, we may
conclude by the pigeonhole principle that there are two separated
but identically colored paths that leave from distinct roots and
lead to leaves corresponding to the same triangle. If we require
in addition that both roots have two incoming edges colored with
the same color, then the same result applies since there are
$2S+1$ trees. By using the definition of $\tilde G$ and the fact
that $G$ cannot be triangle-colored, the construction of two
identically colored distinct cycles can  be continued by
constructing paths from the two nodes corresponding to the
uni-colored triangle to the roots of their corresponding trees. In
this way one can construct two separated but identically colored cycles
in $\tilde G$ that prove that $\tilde G$ cannot be made observable
with two colors and the proof is complete.

\end{proof}
\end{section}

\begin{section}{Conclusion and future work}\label{section-conc}
We have introduced and analyzed a simple notion of observability
in graphs. Observability in graphs is desirable in situations
where autonomous agents move on a network and one want to localize
them; this concept appears in a number of different areas. We have
characterized various forms of observability and have shown how
they can be checked in polynomial time. We have also shown that
the time needed to localize an agent in a graph can be computed in
polynomial time. This time is in the worst case quadratic in the
number of nodes in the graph.  We have also proved that the design
of observable graphs is NP-complete in two distinct situations. We
leave some open questions and problems: Is the problem of making a
graph observable NP-complete if colors are assigned to the edges?
How can one approximate the minimal number of colors? If a graph
can be made observable with a certain number of colors, how can
the colors be assigned so as to minimize the time necessary to
localize the agent?
\end{section}


\begin{thebibliography}{10}
\bibitem{Beal93}
M.-P. Beal.
\newblock {\em Codage symbolique}.
\newblock Masson, 1993.

\bibitem{Crespi05}
V.~Crespi, G.~V. Cybenko, and G.~Jiang.
\newblock {The Theory of Trackability with Applications to Sensor Networks}.
\newblock Technical Report TR2005-555, Dartmouth College, Computer Science,
  Hanover, NH, August 2005.

\bibitem{EphraimM02}
Y.~Ephraim and N.~Merhav.
\newblock Hidden markov processes.
\newblock {\em IEEE Transactions on Information Theory}, 48(6):1518--1569,
  2002.

\bibitem{GJ-computers-igt}
M.~R. Garey and D.~S. Johnson.
\newblock {\em Computers and Intractability; A Guide to the Theory of
  NP-Completeness}.
\newblock W. H. Freeman \& Co., New York, NY, USA, 1990.

\bibitem{JungersProtasovBlondel06}
R.~Jungers, V.~Protasov, and V.D. Blondel.
\newblock Efficient algorithms for deciding the type of growth of products of
  integer matrices.
\newblock submitted to publication.

\bibitem{OzverenWillsky90}
C.~M. \"Ozveren and A.S. Willsky.
\newblock Observability of discrete event dynamic systems.
\newblock {\em IEEE Transactions on automatic and control}, 35:797--806, 1990.

\bibitem{Rabiner90}
L.~R. Rabiner.
\newblock A tutorial on hidden markov models and selected apllications in
  speech recognition.
\newblock In A.~Waibel and K.-F. Lee, editors, {\em Readings in Speech
  Recognition}, pages 267--296. Kaufmann, San Mateo, CA, 1990.

\bibitem{Viterbi67}
A.~J. Viterbi.
\newblock Error bounds for convolutional codes and an asymptotically optimal
  decoding algorithm.
\newblock {\em IEEE Transaction on Information Theory}, IT-13:260--269, 1967.

\end{thebibliography}
\end{document}